\documentclass[10pt, conference]{IEEEtran}

\usepackage[utf8]{inputenc}
\usepackage[T1]{fontenc}
\usepackage{amsmath,amssymb,amsthm}
\usepackage{mathtools}
\usepackage{graphicx}
\usepackage{booktabs}
\usepackage{siunitx}
\usepackage{xcolor}
\usepackage{hyperref}
\usepackage{cleveref}
\usepackage{microtype}

\newcommand{\nmax}{n_{\mathrm{max}}}
\newcommand{\pnet}{p_{\mathrm{net}}}
\newcommand{\pgate}{p_{\mathrm{gate}}}
\newcommand{\ler}{\mathrm{LER}}

\graphicspath{{figures/}}

\title{Design rules for fault-tolerant multi-gate teleportation}

\author{%
  \IEEEauthorblockN{Mathys Rennela}
  \IEEEauthorblockA{Unitary Foundation, France}
}

\date{\today}

\begin{document}

\maketitle

\begin{abstract}

Multi-gate teleportation (MGT) packages $n$ remote gates into a single ebit via a 1-ebit fan-out quantum circuit, saving $n{-}1$ entangled pairs relative to sequential gate teleportation.
The cost is a correlated failure mode: a single network fault propagates through the fan-out tree, injecting a weight-$n$ Pauli error on the target register (weight $n{+}1$ including the control).
We establish a design rule for fault-tolerant packet sizes in rotated surface codes of odd distance~$d$: a rigorous decoder-independent floor $\lfloor d/2 \rfloor$, extended to $\lceil d/2 \rceil$ by a local weight argument confirmed by simulation when the decoder is correlation-aware.
Simulation with PyMatching shows the standard MWPM decoder built from the packet circuit's detector error model (DEM) naturally corrects the correlated error: at $n=\lfloor d/2\rfloor$ the packet matches or surpasses the per-link sequential LER for moderate-to-high $\gamma$, with the crossover $\gamma^\star$ decreasing as $d$ grows (from $\gamma^\star{\approx}17$ at $d{=}5$ to $\gamma^\star{<}1$ at $d{=}9$, extrapolated), whilst reducing the entanglement cost from $n$ ebits to~$1$.
Packetisation wins when the network is the bottleneck ($\gamma \gg 1$); at $\gamma \approx 1$ the packet is marginally worse ($\leq1.13\times$ at $d\in\{5,7\}$, reversing by $d{=}9$) as the $n{-}1$ extra local fan-out gates offset the network savings.
No custom decoding algorithm is required: the standard MWPM decoder, built from the packet circuit's DEM, already encodes the correlation.
\end{abstract}

\section{Introduction}
\label{sec:intro}

Distributed quantum computing (DQC) promises to scale the computational power of quantum processors by interconnecting them. A basic paradigm of DQC is to partition a monolithic circuit into modules which communicate via remote gates, typically implemented by teleportation over an ebit (most often, a shared Bell pair)~\cite{gottesman1999,ejpp}. For a circuit requiring $n$ non-local CNOTs between the same pair of modules, sequential teleportation consumes $n$ ebits.
Multi-gate teleportation (MGT) collapses this to a single ebit by packaging the $n$ gates into a \emph{packet}: one non-local Bell pair drives a local CNOT fan-out tree that distributes the control to all $n$ targets~\cite{andres2019}. The resource savings are clear ($n{-}1$ ebits saved), but the noise profile is not. A single failure on the Bell pair propagates through the fan-out tree, injecting a weight-$n$ correlated Pauli error on the target register (weight $n{+}1$ including the control) rather than $n$ independent weight-1 errors. This correlated failure raises a basic question: \emph{when do the ebit savings survive fault-tolerant quantum error correction?}

The decisive parameter is the network-to-gate noise ratio $\gamma = \pnet/\pgate$.
When $\gamma \gtrsim 1$, the network error dominates and the correlated weight-$n$ error becomes the bottleneck.
Whether an MGT packet remains fault-tolerant, and whether a decoder can exploit the known correlation structure, depends on the interplay between $\gamma$, the code distance $d$, and the packet size $n$.

\subsection*{Related work}

Recent works establish noise thresholds for distributed surface and qLDPC codes~\cite{shaw2026,stack2025,chandra2026,benchasattabuse2026,kaur2026,marton2025,jacinto2026}, including Shalby et al.~\cite{shalby} who use the same $\gamma$ parameter for surface code teleportation interfaces. All address \emph{single}-gate teleportation under \emph{independent} per-link noise; none models a single gate failure injecting a weight-$n$ correlated error. On the resource side, Loke~\cite{loke2026} showed that GHZ-based fan-out reduces $O(n^2)$ non-local resources to $O(n)$, and compiler frameworks~\cite{gragera2026} minimise non-local gate counts as a communication-cost proxy, but neither accounts for the correlated noise that fan-out introduces. This paper connects the two threads, accounting for noise and communication cost simultaneously.

\subsection*{Contributions}
\label{sec:contributions_summary}

(\textit{i}) A single network failure in a 1-ebit fan-out MGT packet injects a weight-$n$ Pauli error on the target register (contiguous in the fan-out ordering), forced by the fan-out bottleneck and independent of the teleportation interface (\cref{sec:error_model}). (\textit{ii}) The MWPM failure threshold at weight $\lceil d/2 \rceil$~\cite{fowler2013,bravyi2013} gives a maximum safe packet size $\lceil d/2 \rceil$ with a correlation-aware decoder, or $\lfloor d/2 \rfloor$ without (\cref{sec:design_rule}); the $\lceil d/2 \rceil$ bound is supported by a local weight argument and confirmed by simulation at $d \in \{5, 7, 9\}$. (\textit{iii}) Detector error model (DEM) level simulation with Stim~\cite{stim} and PyMatching~\cite{pymatching} under a simplified circuit-level depolarising model (\cref{sec:packet_noise}) shows the standard MWPM decoder built from the packet circuit's DEM closes the naive-decoder gap and extends the correctable packet size by one, matching or surpassing the per-link sequential baseline at moderate-to-high $\gamma$ for $d \in \{5, 7, 9\}$, with the advantage growing with $\gamma$ and $d$.
The per-link model (each ebit an independent channel use at rate $\pnet$) is physically motivated: the network error rate is a property of each link, not of the total entanglement budget.
No custom decoding algorithm is required: the natural DEM already encodes the packet structure.

\paragraph*{Assumptions} Simplified circuit-level depolarising noise (gate noise only); use of the Eisert--Jacobs--Papadopoulos--Plenio (EJPP) protocol~\cite{ejpp} on rotated surface codes for quantitative claims. Qualitative conclusions are interface-independent by structural argument (\cref{sec:error_model}); quantitative results use MWPM decoding and the per-link noise model (each ebit an independent channel use at rate $\pnet$).
\section{Error Model}
\label{sec:error_model}

\begin{figure}
  \centering
  \includegraphics[width=0.85\columnwidth]{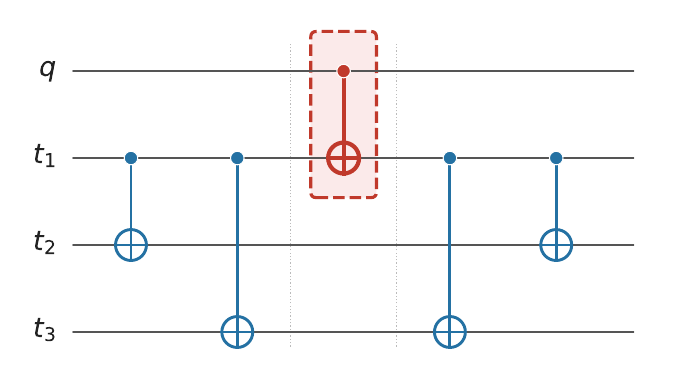}
  \caption{Fan-out decomposition of a same-control MGT packet of 3 CNOTs.  A single non-local CNOT$(q \to t_1)$ (red) is the only operation that crosses the network; all other gates are local fan-out CNOTs (blue).  The total entanglement cost is one ebit, independent of~$n$.}
  \label{fig:fanout_circuit}
\end{figure}

Consider a sequence of consecutive CNOT gates, sharing the same control qubit.
A same-control multi-gate teleportation (MGT) packet is formed by a single non-local CNOT$(q \to t_1)$, preceded and followed by local fan-out CNOTs for $i=2,\ldots,n$, as shown in~\cref{fig:fanout_circuit}.
The non-local CNOT can for example be executed via the EJPP protocol~\cite{ejpp}, which consumes one ebit of entanglement and two rounds of classical communication.
The total entanglement cost of an MGT packet is therefore one ebit, independent of the number of targets $n$.

An $X$ error on $q$ before the non-local CNOT$(q \to t_1)$ propagates as $X_q \to X_q X_{t_1}$, and each subsequent local fan-out CNOT$(t_1 \to t_k)$ copies $X_{t_1}$ onto $t_k$. The result is a weight-$(n{+}1)$ error $X_q X_{t_1} \cdots X_{t_n}$; restricting to the target register, this is a contiguous weight-$n$ $X$ string. An $X$ error on $t_1$ (between the non-local step and the fan-out) produces the same weight-$n$ target string without $q$. Faults on downstream targets $t_k$ ($k > 1$) remain localised. Thus a single network failure injects a weight-$n$ correlated $X$ error; that is, an all-or-nothing event that mimics a logical error chain on the logical $\bar{X}$ operator.

Under depolarising network noise with total rate $p_\text{net}$ (each Pauli fault occurring at rate $p_\text{net}/3$), the dangerous errors on control $q$ include both $X_q$ and $Y_q = iX_q Z_q$: the $X$ component of $Y$ propagates identically through CNOT$(q \to t_1)$ to produce the same weight-$n$ target string, whilst the $Z$ component commutes forward and remains on $q$ alone (see below). Therefore, the effective rate for producing the weight-$n$ $X$ string is $2p_\text{net}/3$ rather than $p_\text{net}/3$.
Simulations in \cref{sec:packet_noise} inject the correlated error at the full $\pnet$ rather than $2\pnet/3$, making reported packet LERs conservative.

$Z$ errors behave asymmetrically and remain secondary: a $Z$ error on $q$ commutes through the non-local CNOT and stays localised, whilst a $Z$ error on a downstream target $t_k$ propagates backward as a weight-$2$ localised error $Z_{t_1} Z_{t_k}$. No $Z$ error produces a weight-$n$ string. The same-target MGT packet (dual configuration) has the opposite behaviour: a $Z$ error on the target produces a weight-$n$ $Z$ string on the controls, whilst $X$ errors remain benign. Without loss of generality, the remainder of this paper treats same-control MGT.

The mechanism extends beyond CNOT to all controlled-Clifford gates (the Clifford property preserves weight and contiguity under conjugation) and is forced by the 1-ebit fan-out bottleneck, making it interface-independent; quantitative claims use the EJPP protocol~\cite{ejpp} but other interconnects may carry additional overhead~\cite{shalby}.

\paragraph*{From fan-out to surface code DEM}
For the design rule and simulations we instantiate the $n$ targets as data qubits of a rotated surface code of distance~$d$, placed on the minimum-weight logical $\bar X$ chain (the $x{=}1$ column). The weight-$n$ packet error is then a contiguous sub-chain of $\bar X$, which is precisely the degenerate-syndrome configuration analysed in \cref{sec:design_rule}: the packet fault and its complement along $\bar X$ produce identical syndromes. The DEM is built from the standard surface code memory circuit with the packet fault injected at round~0 on these $n$ data qubits, so the correlated error appears as a single \texttt{CORRELATED\_ERROR} edge in the matching graph.

\section{Analytic Design Rule}
\label{sec:design_rule}

We established that a single quantum link failure in an MGT packet produces a weight-$n$ correlated error, but how large can a packet be before this correlated error breaks fault tolerance?
We derive the answer from the correctability condition~\cite{knill1997} and the MWPM degenerate syndrome failure mechanism, leading to a design rule that depends on whether the decoder is built from the packet circuit's DEM.

\subsection*{Decoder-independent correctability bound}

A quantum code of distance $d$ can correct any error of weight at most
$\lfloor (d-1)/2 \rfloor = \lfloor d/2 \rfloor$ for odd $d$ (as used
throughout)~\cite{knill1997}. 
For two weight-$n$ errors $E_a, E_b$ with $n \leq \lfloor d/2 \rfloor$,
the product $E_a^\dagger E_b$ has weight at most $2\lfloor d/2 \rfloor = d-1 < d$
and therefore cannot be a nontrivial logical operator.
This bound is decoder-independent and geometry-independent:
a packet of size $n \leq \lfloor d/2 \rfloor$ is correctable regardless of
where the target qubits are located.

\subsection*{Naive decoder: failure at $n = \lceil d/2 \rceil$}

At $n = \lceil d/2 \rceil$, the failure mechanism changes as follows.
Place a weight-$n$ $X$ error $E = \prod_{j \in S} X_{q_j}$ on a subset $S$
(with $|S| = n$; contiguous in the fan-out error model) of
the minimum-weight logical $\bar{X}$ operator~$L$.
Define its complement $E' = \prod_{j \in L \setminus S} X_{q_j}$ on
the remaining $d - \lceil d/2 \rceil = \lfloor d/2 \rfloor$ qubits.
Since $E \cdot E' = \bar{X}$ commutes with all stabilisers, $E$ and $E'$
produce a degenerate pair of identical syndromes.
Because $|E| = \lceil d/2 \rceil$ exceeds the correctable weight $\lfloor d/2 \rfloor$
whilst $|E'| = \lfloor d/2 \rfloor$ does not, a \emph{naive} MWPM decoder
(built from the base DEM, without the correlated-error edge) selects~$E'$ as the minimum-weight correction.
The residual $E \cdot E' = \bar{X}$ is a logical failure: an error of weight $\lceil d/2 \rceil$
is uncorrectable whenever its complement along a logical operator has weight $\lfloor d/2 \rfloor$.
This mechanism is purely algebraic: it depends on the stabiliser structure,
not on the matching-graph geometry.

\subsection*{Correlation-aware decoder: extension to $n = \lceil d/2 \rceil$}

The failure mechanism above assumes the decoder treats the weight-$n$ correlated error as $n$ independent gate-error edges.
A correlation-aware decoder (which includes the correlated error as a single \texttt{CORRELATED\_ERROR} edge) assigns it a weight of $-\ln \pnet$ rather than $n \cdot (-\ln \pgate)$.
At $n = \lceil d/2 \rceil$, the complement $E'$ has weight $\lfloor d/2 \rfloor \cdot (-\ln \pgate)$ in the matching graph.
Whenever $\gamma \geq 1$, the single packet edge is cheaper:
$-\ln \pnet \leq -\ln \pgate < \lfloor d/2 \rfloor \cdot (-\ln \pgate)$.
This comparison is local: between the degenerate pair $\{E, E'\}$ it favours the \emph{correct} correction~$E$, leaving no residual, but it does not exclude a third error class yielding an even cheaper matching for the same syndrome.
This weight comparison is heuristic, not a proof.
Computational confirmation at $d{=}5,7,9$ supports it: at $n = \lceil d/2 \rceil$ the correlation-aware LER stays within $\sim\!2$--$3\times$ of the gate-noise floor ($\sim 10^{-4}$ at $d{=}5$, $\sim 10^{-5}$ at $d{=}7$, $\sim 3$--$5\times10^{-7}$ at $d{=}9$) across $\gamma \in \{1, 10, 100\}$, whilst the naive decoder fails ($\ler \to 0.01$--$0.1$). The naive decoder is also \emph{degraded} below $n_{\max}$ at high $\gamma$ (e.g.\ $n{=}2$, $d{=}5$, $\gamma{=}100$: naive LER $\sim 1.3\times10^{-3}$, $\approx 16\times$ the correlation-aware value), since it lacks the correlated-error edge.

\begin{figure}[t]
  \centering
  \includegraphics[width=\columnwidth, height=0.48\textheight, keepaspectratio]{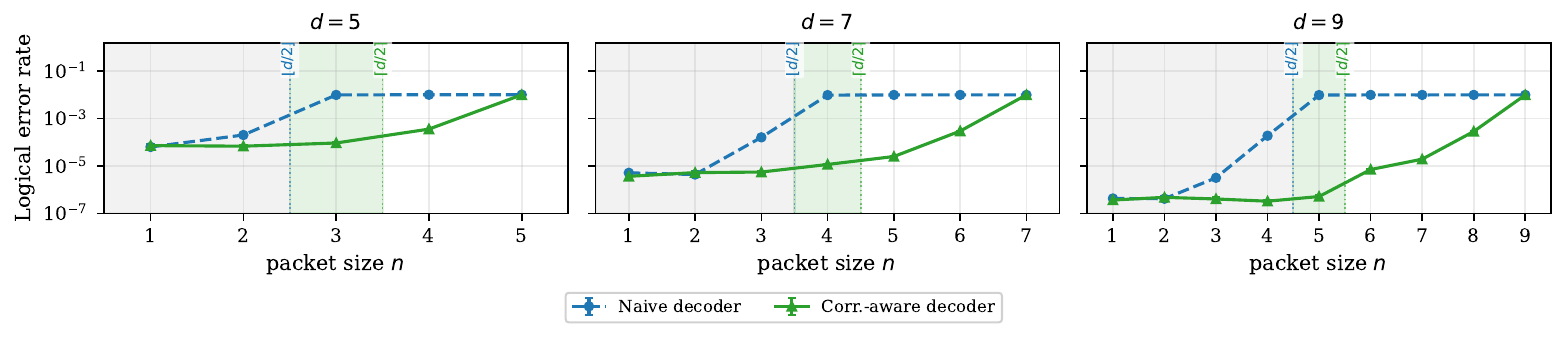}
  \caption{Naive vs.\ correlation-aware LER vs.\ packet size $n$ at $\gamma = 10$, $d \in \{5,7,9\}$. Blue dashed: naive decoder; green solid: correlation-aware decoder. Grey shading: both correctable; green shading: correlation-aware-only extension. At $n{=}\lceil d/2 \rceil$ the correlation-aware LER is within $\sim\!2$--$3\times$ of the gate-noise floor whilst the naive decoder fails. $d{=}9$ uses $10^8$ shots; $d{=}5,7$ use $10^6,10^7$. $\pgate = 10^{-3}$.}
  \label{fig:nmax_validation}
\end{figure}

\Cref{fig:nmax_validation} sweeps $n$ from $1$ to $d$ for $d \in \{5, 7, 9\}$ at $\gamma = 10$. The naive decoder cliffs at $n = \lceil d/2 \rceil$; the correlation-aware decoder holds near the floor through $\lceil d/2 \rceil$, degrades monotonically above it (to $\sim\!10^{-4}$ at $n{=}d{-}1$), and converges with the naive decoder at $\ler \approx \pnet$ for $n{=}d$ (zero-syndrome limit).

\subsection*{Design rule}

Combining the bounds gives two design rules for MGT packet size:
\begin{equation}
  \nmax^{\text{naive}}(d) = \left\lfloor \frac{d}{2} \right\rfloor,
  \qquad
  \nmax^{\text{corr}}(d) = \left\lceil \frac{d}{2} \right\rceil
  \label{eq:nmax}
\end{equation}
for odd $d$, where $\nmax^{\text{naive}}$ applies to a decoder built from the base DEM and $\nmax^{\text{corr}}$ to a correlation-aware decoder.
For $d = 5, 7, 9, 11$, this gives $\nmax^{\text{corr}} = 3, 4, 5, 6$.
The decoder-independent correctability floor is $\lfloor d/2 \rfloor$ in both cases; exceeding $\nmax^{\text{corr}}$ forces a degenerate syndrome with no lower-weight complement.

The upper bound requires worst-case placement: all $n$ targets on $\bar{X}$ so the complement has weight $\lfloor d/2\rfloor$; a limited $d{=}5$ sweep is consistent with on-chain being canonical, but this is not proven in general. Only one Pauli type per configuration is dangerous (weight-$n$ $X$ for same-control); the benign type's worst case is weight-$2$, correctable for all $d \geq 5$.

\section{Packet Fault-Tolerance Performance}
\label{sec:packet_noise}

We compare the packet model (single $\texttt{CORRELATED\_ERROR}(\pnet)$, weight~$n$) against the per-link sequential baseline ($n$~independent $\texttt{X\_ERROR}(\pnet)$ per link), where $\pnet$ is the per-link rate and each ebit is an independent channel use (\cref{sec:error_model}); $\gamma = \pnet/\pgate$ parametrises the noise ratio; $\pgate = 10^{-3}$ throughout.

When the decoder does not account for the correlated error, the relative cost grows
with code distance at moderate-to-high $\gamma$ (see $\ler_{\mathrm{base}}$ in \cref{tab:architecture}).

However, this gap is an artefact of the decoder's DEM, not of the packet architecture.
The packet circuit's DEM naturally includes the correlated error as a single edge.
Building a standard MWPM decoder from this DEM closes the gap (the construction requires only weight-capping and edge-ordering workarounds for the PyMatching implementation, not a bespoke matching algorithm):\footnote{PyMatching's \texttt{merge\_strategy="independent"} drops the observable attribution of a boundary edge added after a same-node gate-noise edge; fault-carrying edges must be inserted first.}
\cref{tab:architecture} shows that the correlation-aware packet LER
($\ler_{\mathrm{pkt}}$) drops to the gate-noise floor, $\gamma$-independent,
because the decoder absorbs the single packet fault.
Against the per-link sequential baseline ($n$ independent faults at rate
$\pnet$ each), the packet matches or surpasses sequential at moderate-to-high
$\gamma$, with the advantage growing with both $\gamma$ and $d$.
At low $\gamma$ ($\gamma \approx 1$), the packet is slightly worse at
$d \in \{5,7\}$ (ratio $\leq 1.13$) because the $n{-}1$ extra local fan-out
CNOTs contribute gate noise without decisive network savings; this reverses
by $d{=}9$ (ratio $0.27$).
At high $\gamma$, the sequential baseline accumulates $n$ independent
network faults whilst the correlation-aware decoder absorbs the single packet
fault at the gate-noise floor.
We define the crossover $\gamma^\star$ as the value of $\gamma$ where pkt/seq
$=1$ in \cref{tab:architecture}: log-log interpolation gives
$\gamma^\star \approx 17$ at $d{=}5$ and $\approx 6.3$ at $d{=}7$, and
(extrapolated past the lowest tested $\gamma{=}1$) $\gamma^\star < 1$ at
$d{=}9$, a monotonic decrease driven by the sequential baseline's
$n{=}\lceil d/2 \rceil$ independent network-fault chances versus the
packet's one, a gap that widens with $d$.


\begin{table}
  \caption{Packet LER: correlation-agnostic DEM ($\ler_{\mathrm{base}}$), correlation-aware DEM ($\ler_{\mathrm{pkt}}$), per-link sequential baseline ($\ler_{\mathrm{seq}}$). $\pgate\!=\!10^{-3}$; $10^6$--$10^8$ shots; $\pm 1\sigma$ Poisson; $95\%$ upper bound for zero counts.}
  \label{tab:architecture}
  \centering
  \resizebox{\columnwidth}{!}{%
  \small
  \begin{tabular}{@{} ccc ccc c @{}}    \toprule
    $d$ & $n$ & $\gamma$ & $\ler_{\mathrm{base}}$ &
    $\ler_{\mathrm{pkt}}$ & $\ler_{\mathrm{seq}}$ & pkt/seq \\
    \midrule
    5 & 2 & 1   & $(9.20 \pm 0.96)\times10^{-5}$ & $(7.11 \pm 0.08)\times10^{-5}$ & $(6.91 \pm 0.08)\times10^{-5}$ & 1.03 \\
    5 & 2 & 10  & $(1.86 \pm 0.14)\times10^{-4}$ & $(7.45 \pm 0.09)\times10^{-5}$ & $(7.02 \pm 0.08)\times10^{-5}$ & 1.06 \\
    5 & 2 & 30  & $(4.08 \pm 0.20)\times10^{-4}$ & $(7.67 \pm 0.09)\times10^{-5}$ & $(8.18 \pm 0.09)\times10^{-5}$ & 0.94 \\
    5 & 2 & 100 & $(1.27 \pm 0.04)\times10^{-3}$ & $(7.95 \pm 0.09)\times10^{-5}$ & $(1.92 \pm 0.01)\times10^{-4}$ & 0.41 \\
    \midrule
    7 & 3 & 1   & $(2.05 \pm 0.05)\times10^{-5}$ & $(5.91 \pm 0.24)\times10^{-6}$ & $(5.21 \pm 0.23)\times10^{-6}$ & 1.13 \\
    7 & 3 & 10  & $(1.66 \pm 0.01)\times10^{-4}$ & $(5.33 \pm 0.23)\times10^{-6}$ & $(5.50 \pm 0.23)\times10^{-6}$ & 0.97 \\
    7 & 3 & 30  & $(4.85 \pm 0.02)\times10^{-4}$ & $(5.20 \pm 0.23)\times10^{-6}$ & $(8.0 \pm 2.8)\times10^{-6}$ & 0.65 \\
    7 & 3 & 100 & $(1.604 \pm 0.004)\times10^{-3}$ & $(5.86 \pm 0.24)\times10^{-6}$ & $(3.50 \pm 0.59)\times10^{-5}$ & 0.17 \\
    \midrule
    9 & 4 & 1   & $(1.89 \pm 0.04)\times10^{-5}$ & $(5.30 \pm 0.73)\times10^{-7}$ & $(2.0 \pm 1.4)\times10^{-6}$ & 0.27 \\
    9 & 4 & 10  & $(1.93 \pm 0.01)\times10^{-4}$ & $(4.20 \pm 0.65)\times10^{-7}$ & $(1.00 \pm 1.00)\times10^{-6}$ & 0.42 \\
    9 & 4 & 30  & $(5.75 \pm 0.02)\times10^{-4}$ & $(4.00 \pm 0.63)\times10^{-7}$ & $<\!3\!\times\!10^{-6}$ & --- \\
    9 & 4 & 100 & $(1.913 \pm 0.004)\times10^{-3}$ & $(5.00 \pm 0.71)\times10^{-7}$ & $(3.0 \pm 1.7)\times10^{-6}$ & 0.12 \\
    \midrule
    \multicolumn{7}{l}{\small Design-rule validation: $n = \lceil d/2 \rceil$} \\
    \midrule
    5 & 3 & 1   & $(1.082 \pm 0.033)\times10^{-3}$ & $(8.3 \pm 0.9)\times10^{-5}$ & --- & --- \\
    5 & 3 & 10  & $(1.002 \pm 0.100)\times10^{-2}$ & $(9.4 \pm 1.0)\times10^{-5}$ & --- & --- \\
    5 & 3 & 100 & $(9.95 \pm 0.30)\times10^{-2}$  & $(1.89 \pm 0.14)\times10^{-4}$ & --- & --- \\
    \midrule
    7 & 4 & 1   & $(9.84 \pm 0.10)\times10^{-4}$ & $(8.0 \pm 0.9)\times10^{-6}$ & --- & --- \\
    7 & 4 & 10  & $(9.82 \pm 0.31)\times10^{-3}$ & $(1.17 \pm 0.11)\times10^{-5}$ & --- & --- \\
    7 & 4 & 100 & $(9.88 \pm 0.10)\times10^{-2}$ & $(1.70 \pm 0.13)\times10^{-5}$ & --- & --- \\
    \bottomrule
  \end{tabular}%
  }
\end{table}

All sweeps use pinned random seeds for reproducibility.

\section{Conclusion}
\label{sec:conclusion}

Multi-gate teleportation~\cite{andres2019} saves $n{-}1$ ebits by packaging $n$ remote CNOTs behind a single non-local crossing, at the cost of a correlated failure mode: a single network fault produces a weight-$n$ Pauli error on the target register (\cref{sec:error_model}). 
Two design rules govern fault-tolerant MGT for rotated surface codes. \emph{First}, the packet size must satisfy $n \leq \lceil d/2 \rceil$ with a correlation-aware decoder, or the stricter $n \leq \lfloor d/2 \rfloor$ with a naive decoder (\cref{sec:design_rule}); the correctability floor $\lfloor d/2 \rfloor$ is decoder-independent, with the only proven hard limit at $n=d$. 
\emph{Second}, the decoder must be built from the packet circuit's DEM, which encodes the correlated error as a single edge (\cref{sec:packet_noise}). With it, the packet matches or surpasses the per-link sequential baseline at moderate-to-high $\gamma$, and decisively outperforms at high $\gamma$ or high $d$, where the sequential baseline accumulates $n$ independent network faults whilst the decoder absorbs the single packet fault.

The correlated error originates from the 1-ebit bottleneck, so this analysis is interface-independent (\cref{sec:error_model}). For packets exceeding $\nmax^{\text{corr}}$, a $k$-ary fan-out tree restores fault tolerance at the cost of $k$ ebits: each of $k$ bridge qubits fans out to a subgroup of $\approx n/k$ targets, reducing the worst-case error weight from $n$ to $\approx n/k$, with $k_{\min} = \lceil n / \lceil d/2 \rceil \rceil$ interpolating between 1-ebit fan-out ($k{=}1$) and fully sequential teleportation ($k{=}n$).

\paragraph*{Limitations and future work}
The simplified model (gate noise only) underestimates absolute LER by $3$--$5\times$ but preserves the packet-vs-sequential ranking since omitted channels enter both DEMs identically. 
Photonic CNOT fidelities of $\sim 87\%$~\cite{chang2025} and entanglement error rates of $\sim 5\%$~\cite{jiang2007} give $\gamma \in [50, 100]$, where packetisation already outperforms; our DEM injects at full $\pnet$ rather than $2\pnet/3$ (\cref{sec:error_model}), making LERs conservative. 
Simulation places the packet error at round~0 (maximal syndrome information); a final round injection would be harder. 
Quantitative claims use EJPP~\cite{ejpp} with rotated surface codes and MWPM decoding; qualitative conclusions extend by structural argument to any surface code and minimum-weight decoder. 
The $\nmax$ bound applies to any CSS code, but evidence is limited to rotated surface codes; extending to qLDPC codes~\cite{shaw2026,chandra2026}, biased noise, full circuit-level noise, and end-to-end benchmarks via compiler integration~\cite{tourgross,bandini2026,mengoni2025,liu2026,burt2025} is deferred to future work.

\paragraph*{Acknowledgements} This work has been supported by the Mozilla Foundation. AI assistance (GLM-5.2, Mimo-V2.5) was used for copy-editing, proofreading, and formatting, and helped with numerical data generation and analysis. All scientific content and conclusions are the sole responsibility of the author.

\paragraph*{Data availability} The detector error models (DEMs) underlying all simulations in this work are publicly available on Zenodo at \href{https://doi.org/10.5281/zenodo.21110299}{10.5281/zenodo.21110299}.

\bibliographystyle{IEEEtran}
\bibliography{references}

\end{document}